\documentclass{ws-ijmpb}
\usepackage{color}

\begin{document}

\markboth{M.Sato, H. Shima and K. Iiboshi}
{Core-tube morphology of multiwall carbon nanotubes}

%
\catchline{}{}{}{}{}
%

\title{CORE-TUBE MORPHOLOGY OF MULTIWALL CARBON NANOTUBES}

\author{MOTOHIRO SATO}

\address{Department of Socio-Environmental Engineering, Graduate School of Engineering,\\
Hokkaido University, Sapporo 060-8628, Japan\\
tayu@eng.hokudai.ac.jp}

\author{HIROYUKI SHIMA}

\address{Department of Applied Physics, Graduate School of Engineering,\\
Hokkaido University, 060-8628 Japan\\
shima@eng.hokudai.ac.jp}

\author{KOHTAROH IIBOSHI}

\address{Department of Socio-Environmental Engineering, Graduate School of Engineering,\\
Hokkaido University, Sapporo 060-8628, Japan\\
jojo.uomo@hotmail.co.jp}

\maketitle

\begin{history}
\received{\today}
\revised{Day Month Year}
\end{history}

\begin{abstract}
The present paper investigates the cross-sectional morphology of 
Multiwalled Carbon Nanotubes (MWNTs) restrained radially and circumferentially 
by an infinite surrounding elastic medium, 
subjected to uniform external hydrostatic pressure. 
In this study, a two-dimensional plane strain model is developed, 
assuming no variation of load and deformation along the tube axis. 
We find some characteristic cross-sectional shapes from the elastic buckling analysis. 
The effect of the surrounded elastic medium on the cross-sectional shape 
which occurs due to pressure buckling is focused on 
by the comparison with the shape for no elastic medium case in our discussion. 
It is suggested that in no embedded elastic medium cases, 
the cross-sectional shapes of inner tubes maintain circle or oval; 
on the other hand, an embedded medium may cause inner tube corrugation modes 
especially when the number of shells for MWNTs is small.

\end{abstract}


\section{Introduction}

Multiwalled carbon nanotubes (MWNTs) consist of graphene sheets wrapped up to form
a series of concentric cylindrical shells\cite{Popov2004,IMM}.
The number of shells $N$ in a MWNT ranges from 2 to more than 100, and they mutually interact
via van der Waals (vdW) forces.
Of interest is the feature that internal structures of MWNTs,
like differences in diameters and chiralities
between consecutive shells,
are responsible for electronic\cite{Park1999,Gomez2006,Cai2006,Monteverde2006,Nishio2008,Barboza2008,Ren2009}
and optical\cite{Venkateswaran1999,Deacon2006,Lebedkin2006,Longhurst2007,Yao2008,Chang2008} 
properties of the systems. 
Therefore, better understandings of their cross-sectional deformation driven by external forces
would be important for developing various MWNT-based applications involving 
sensors\cite{Sinha2006,Zhang2008} and actuators \cite{actuator2008}.
Nevertheless, most of earlier studies have dealt only with single-walled carbon nanotubes,
in which simple radial collapse under pressures of the order of a few GPa was observed \cite{J_Tang2000,Zhang2006}.

Intuitively, the core-shell structure of a MWNT is expected to enhance its radial stiffness,
as restoring forces exerted on pairs of neighboring shells
buffer against cross-sectional deformation.
This conjecture is, nevertheless, not obvious when $N\gg 1$.
For the latter case, outside shells have large diameters that lead to mechanical instability,
whereas inside shells keep the radial rigidity due to their small diameters.
Such the discrepancy in the radial stiffness between outside and inside shells 
implies a new class of cross-sectional deformation induced by hydrostatic pressure.
Another interesting issue is the radial and circumrerential stiffness of MWNTs embedded in an elastic medium.
When applying an external force to the host medium of elastic composite embedding MWNTs,
then a restoring force acts on the outermost surface of MWNTs.
This restoring force will tend to remain the circular geometry of 
the outermost shell of each dispersed MWNT; 
this is in contrast with the previous case where the outermost shell become softened.
Exploring the cross-sectional morphology of embedded MWNTs will provide useful information for developing
nanofluidic \cite{Majumder2005,Noy2007,Whitby2007,Khosravian2008,Mattia2008}
or nanoelectrochemical \cite{Frackowiak2001,Tasis2006} devices
whose performance depend crucially on 
the geometry of the inner hollow cavity of nanotubes.

In this article, we analytically demonstrate the characteristic 
cross-sectional morphology of MWNTs with and without embedded elastic medium 
under external pressure. 
A two-dimensional plane strain model is developed, assuming no variation of 
load and deformation along the tube axis. 
It has been found that the existence of the surrounded elastic medium 
crucially affect the cross-sectional shape under buckling state. 
Some examples on the interesting cross-sectional shapes are introduced 
and brief discussions on these phenomena are also given.

\section{Method}

We have employed the continuum model of MWNTs\cite{Ru,Shen,He2005,NTN,PSSA}
that enables to deduce the stable cross-sectional geometry under pressure.
Though atomistic simulations are realistic and accurate,
the computational cost associated is huge.
This has encouraged the use of continuum models
in various fields to study mechanical properties of MWNTs,
together with discussions on the validity of the models\cite{Sears2004,Huang2006,Peng2008,Lu2009}.

In the continuum model, a MWNT is mapped onto a set of $N$ concentric cylindrical shells
with thickness $h$, wherein the $i$th shell has the radius $r_i$.
The mechanical energy $U$ of a MWNT per unit axial length is written as
\begin{equation}
U = U_D + U_I + \Omega.
\label{eq_01}
\end{equation}
The first term $U_D = \sum_{i=1}^N U_D^{(i)}$ represents the deformation energy,
in which $U_D^{(i)} = (k r_i/2)
\int_{-h/2}^{h/2} \int_0^{2\pi} \tilde{\varepsilon}_i (z_i,\theta)^2 dz_i d\theta$,
$\tilde{\varepsilon}_i$ is the extensional strain 
of a circumferential line element;
$z_i$ is a radial coordinate measured from the centroidal curve of the $i$th shell,
$\theta$ is a circumferential coordinate,
and $k=E/(1-\nu^2)$ with the Young modulus $E = 1$ TPa
and Poisson's ratio $\nu = 0.27$ of MWNTs.
The second term  $U_I$ in Eq.~(\ref{eq_01}) is
the interaction energy of all adjacent pairs of shells described by
$U_I = \sum_{i,j} c_{i,j} r_i \int_0^{2\pi} 
\left( u_i - u_{i+1} \right)^2 d\theta$,
where $u_i$ describes the radial deformation of the $i$th shell,
and the summation is taken for all neighboring pairs of $(i,j)$.
The vdW interaction coefficients $c_{ij}$ are functions of $r_i$ and $r_j$
as proved in Ref.~\refcite{He2005}.
The final term $\Omega$ in Eq.~(\ref{eq_01}) is the negative of the work done by the external pressure $p$
during cross-sectional deformation;
it is expressed as\cite{PSSA}
$\Omega = - p (\pi r_N^2 - S^*)$,
where $S^*$ is the area surrounded by the $N$th shell
after deformation.
Note that the sign of $p$ is assumed to be positive inward.
The stable cross-sectional shape of a MWNT under pressure $p$
is deduced by applying the variation method to $U$ in Eq.~(\ref{eq_01}) 
with respect to $u_i$ and $v_i$ \cite{NTN}.

When the outermost surface of MWNTs is surrounded and 
perfectly bonded by an elastic medium,
the strain energy 
\begin{equation}
U_C = \frac{r_N}{2} \int_0^{2\pi} \left(\sigma_r u_N + \sigma_{\theta} v_N \right)d\theta
\end{equation}
should be added \cite{Shima_unp} in the total mechanical energy given by Eq.~(\ref{eq_01});
here $v_N$ is the deformation of the surface in the circumferential direction,
and $\sigma_r$, $\sigma_{\theta}$ are radial and circumferential normal stresses
in the elastic medium, caused by the deformation of the nanotube
surface, respectively.

\section{Outer-shell corrugation under hydrostatic pressure}

\begin{figure}[ttt]
\begin{center}
\includegraphics[width=3.9cm]{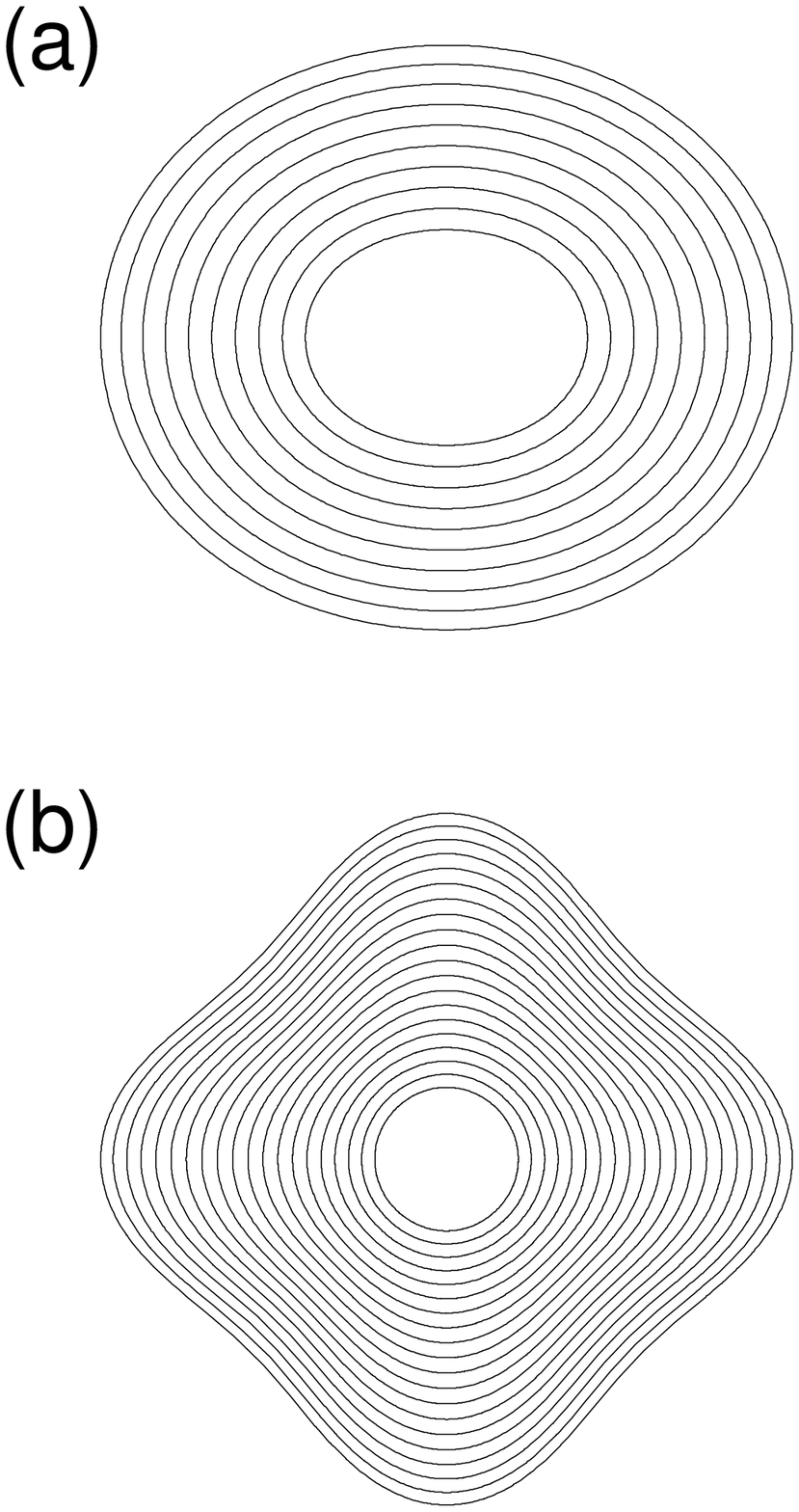}
\hfill
\includegraphics[width=8.5cm]{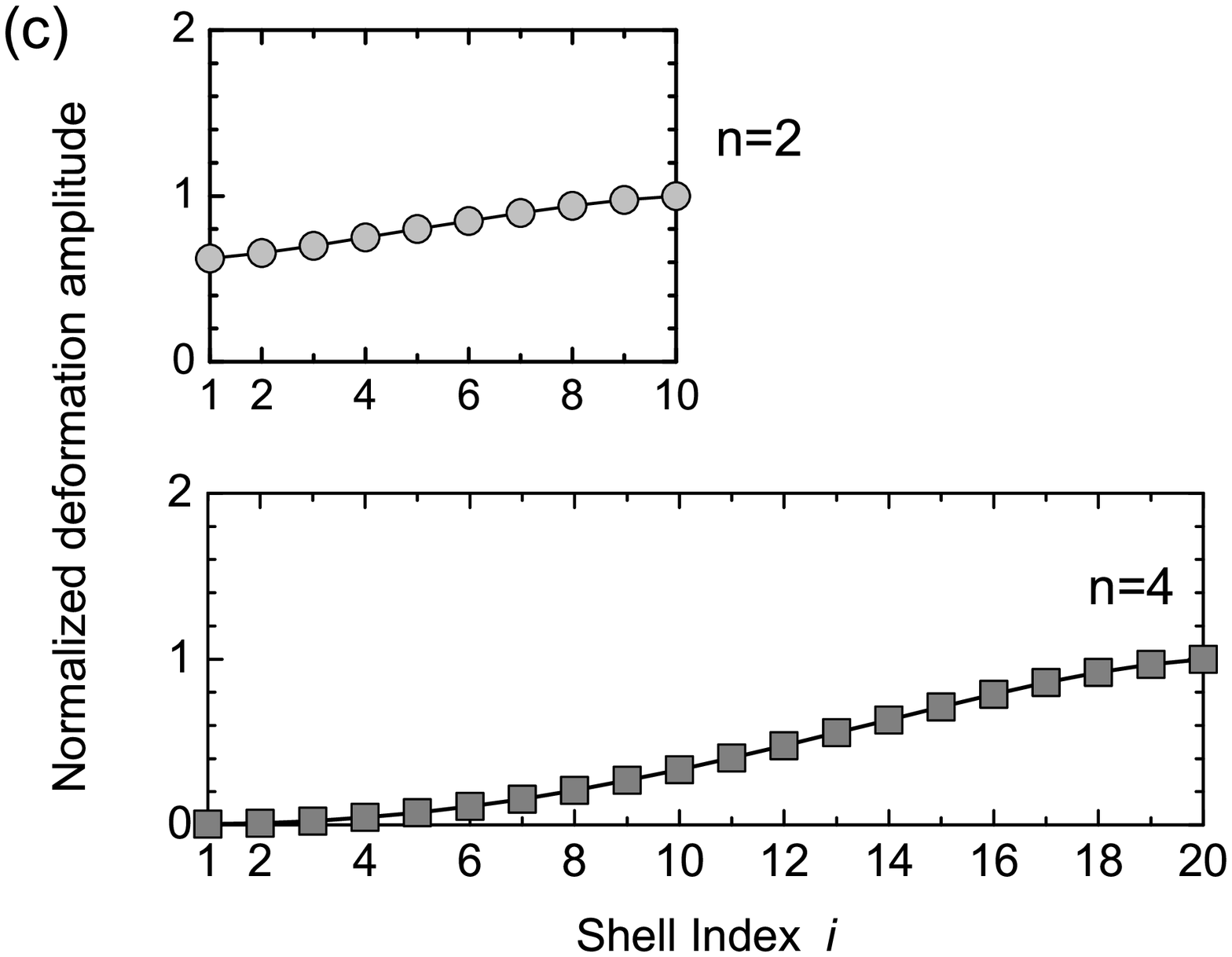}
\end{center}
\caption{(a-b) Cross-sectional views of (a) elliptic $(n=2)$ and (b) corrugated $(n=4)$ 
deformation modes observed for MWNTs with $N=10$ and $N=20$, respectively.
The innermost shell diameter $D=4.0$ nm is fixed.
(c) Relative deformation amplitude of each $i$th concentric shell
subject to the two modes given in (a) and (b).}
\label{fig_01}
\end{figure}

Figure \ref{fig_01} (a) and (b) illustrate elastic buckling modes of MWNTs with $N=10$ and $N=20$
under hydrostatic pressure, respectively; $D\equiv 2r_1 =4.0$ nm is fixed,
and the mode index $n$ indicates the wave number of the deformation mode along the circumference.
In the elliptic mode with $n=2$ shown in Fig.~1 (a),
all concentric shells are almost uniformly deformed
so that consecutive shells are equally spaced even after deformation.
This behavior becomes more clear by plotting the normalized deformation amplitudes
of individual concentric shells as given in the upper plot of Fig.~\ref{fig_01} (c).
By increasing $N$, we obtain qualitatively different results
as shown in Fig.~1(b), where the cross-section yields a corrugation mode of $n=4$.
In the latter mode, outside shells exhibit significant deformation,
while the innermost shell maintains its cylindrical symmetry.

We point out that the persistence of cylindrical symmetry of the innermost shell
will be useful in developing nanotube-based nanofluidic
\cite{Majumder2005,Noy2007,Whitby2007,Khosravian2008}
or nanoelectrochemical devices \cite{Frackowiak2001,Tasis2006},
as both utilize the hollow cavity within the innermost shell.
In fact, manifold class of intercalated molecules
can fill the innermost hollow cavities of nanotubes \cite{Noy2007,Shanavas2009},
which trigger intriguing behaviors distinct from those of 
the corresponding bulk systems \cite{Monthioux2002,CKYang2003,Joseph2008}.
In particular, the innermost shell of MWNTs in the radial corrugation mode
serves as an ideal protective shield for the intercalates,
since it maintains its cylindrical geometry 
even under high external pressure.

\section{Inner-shell corrugation of embedded MWNTs}

\begin{figure}[ttt]
\begin{center}
\includegraphics[width=3.9cm]{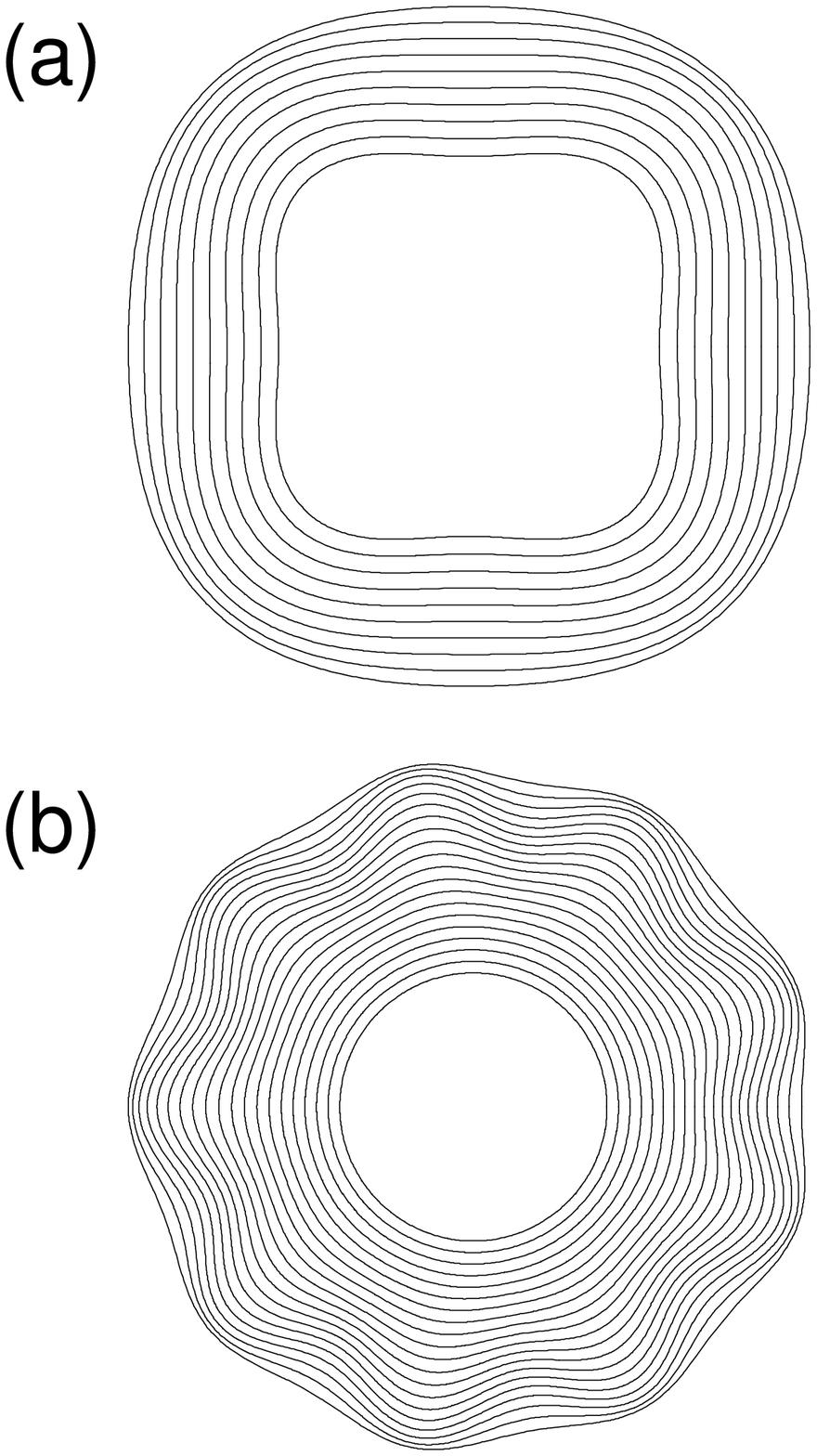}
\hfill
\includegraphics[width=8.5cm]{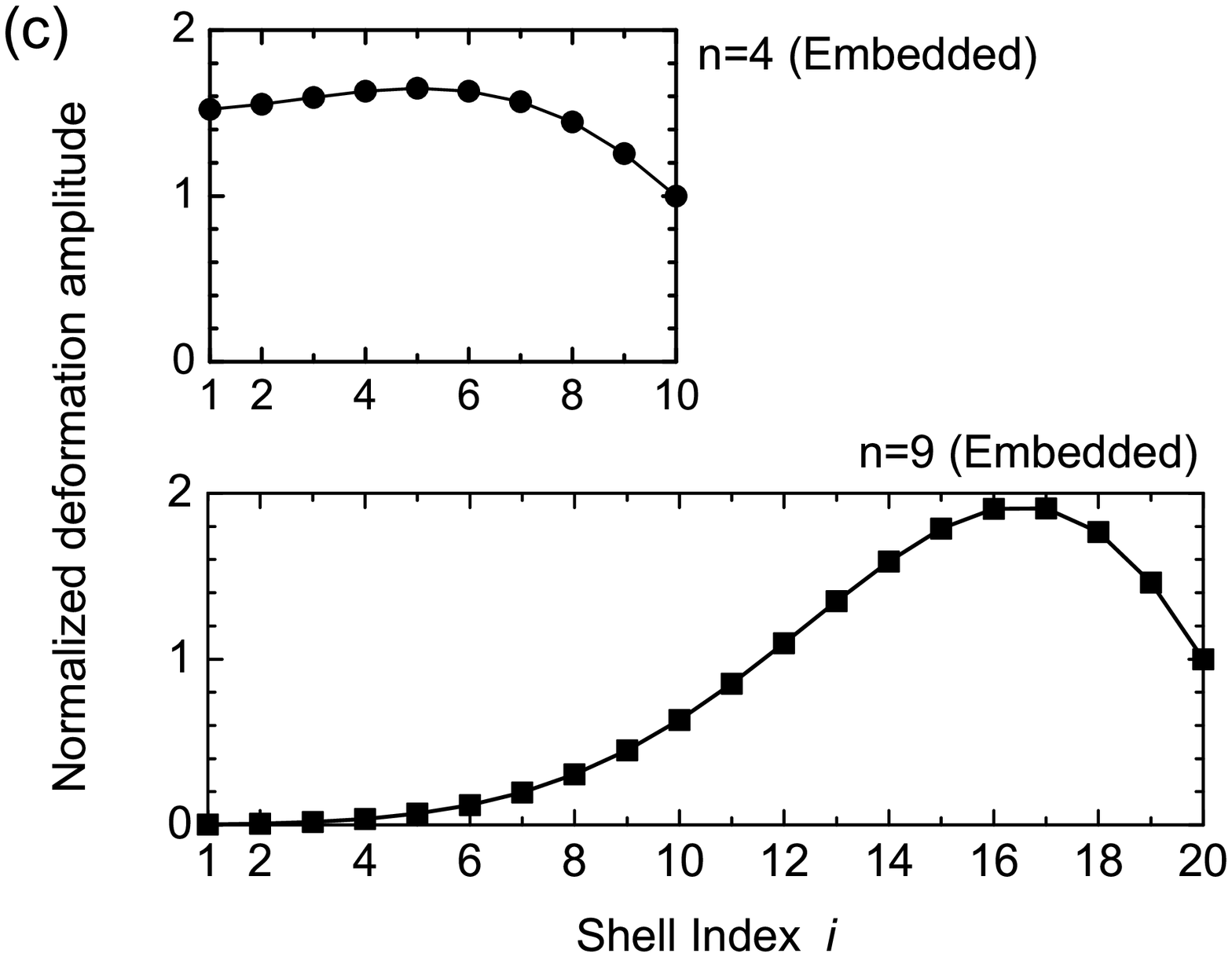}
\end{center}
\caption{(a-b) Two corrugation modes of (a) $n=4$ and (b) $n=9$,
for MWNTs embedded in an elastic medium with $E_M = 100$ GPa and $\nu_M=0.3$.
The number of shells are $N=10$ and $N=20$, respectively,
and $D=8.0$ nm is fixed.
(c) Plots of relative deformation amplitudes.}
\label{fig_02}
\end{figure}

Figure \ref{fig_02} (a) and \ref{fig_02} (b) show
two corrugation modes for MWNTs surrounded by an elastic medium 
with $E = 100$ GPa and $\nu = 0.3$;
Figure \ref{fig_02} (c) indicates plots of relative deformation amplitude for the embedded case. 
From here we can see two characteristic tendencies which is not shown in the case of 
no elastic medium described in the previous section.
First, corrugation mode exhibits in not only outside but also inside shells for a MWNT with $N = 10$.
Second, the corrugation mode that remain its cylindrical symmetry is shown for $N = 20$, however, 
the deformation amplitude of the outermost shell is not always largest. 
This fact says that an embedded elastic medium may cause ``inner tube instability".

\section{Concluding remarks}

It has been found that in radial corrugation modes,
large discrepancy in the deformation amplitudes between the outer and inner shells
perturb the equal spacing between the concentric shells of the original MWNTs.
In no embedded elastic medium cases, the cross-sectional shapes of inner shells 
maintain circle or oval. 
On the other hand, an embedded medium may cause inner shell corrugation modes 
especially when the number of shells for MWNTs is small.
Such the modulation in inter-shell spacing will affect
electronic and vibrational properties of the entire nanotube,
thus triggering a change in its electronic and thermal conductance.

We also conjecture that those pressure-induced changes in physical properties of MWNTs
are of practical use for developing MWNT-based pressure sensors.
It is particularly hoped that our results should be verified by high-pressure experiments
on MWNTs as well as atomic-scale large-scale simulations \cite{Arias,Arroyo,Arroyo2}.

Before closing, it deserves comments on the relevance of our results 
to quantum transport of low-dimensional nanostructures.
It is theoretically known \cite{curve2} that quantum particles moving along a thin curved surface
behave differently from those on a conventional flat plane,
since nonzero geometric curvature results in 
an effective electromagnetic field \cite{ShimaPRB,ShimaPhysicaE,Ono,OnoPhysicaE,Taira}.
In this context, quantum transport of electrons moving in MWNTs under pressure
will exhibit non-trivial surface curvature effects, though details are to be explored.
Intensive studies on the issues mentioned above will shed light on novel MWNT applications
based on cross-sectional deformation.

\section*{Acknowledgements}
We acknowledge Professor T.~Mikami for stimulating discussions.
This study was supported by a Grant-in-Aid for Scientific Research 
on Innovative Areas and the one for Young Scientists (B) from the MEXT, Japan.
M.S and H.S are thankful for financial supports from 
Hokkaido Gas Co., Ltd. and 
Executive Office of Research Strategy in Hokkaido University.
H.S is also grateful for the support from the Kajima Foundation.
A part of the numerical simulations were carried out using
the facilities of the Supercomputer Center, ISSP, University of Tokyo.


\end{document}